\documentclass[12pt,a4paper]{article}
\usepackage{graphicx}
\usepackage{cite}

\makeatletter
\renewcommand{\@biblabel}[1]{}
\makeatother

\begin{document}

\newcommand{\EQ}{Eq.~}
\newcommand{\EQS}{Eqs.~}
\newcommand{\FIG}{Fig.~}
\newcommand{\FIGS}{Figs.~}
\newcommand{\SEC}{Sec.~}
\newcommand{\SECS}{Secs.~}

\setlength{\baselineskip}{0.77cm}

\title{Participation costs
dismiss the advantage of heterogeneous networks
in evolution of cooperation}
\bigskip

\author{Naoki Masuda${}^*$\\
\  \\
\ \\
${}^{1}$
Graduate School of Information Science and Technology,\\
The University of Tokyo,\\
7-3-1 Hongo, Bunkyo, Tokyo 113-8656, Japan
\ \\
$^*$ Author for correspondence (masuda@mist.i.u-tokyo.ac.jp)}
\date{}
\maketitle


\newpage

\begin{abstract}
\setlength{\baselineskip}{0.77cm} 
Real social interactions occur on
networks in which each individual is connected to some, but not all,
of others.  In social dilemma games with a fixed population size,
heterogeneity in the number of contacts per player is known to promote
evolution of cooperation. 
Under a common assumption of
positively biased payoff structure, well-connected
players earn much by playing frequently, and
cooperation once
adopted by well-connected players is unbeatable and spreads to
others.  However, maintaining a social contact can be costly,
which would prevent local payoffs from
being positively biased. In replicator-type evolutionary
dynamics, it is shown that
even a relatively small participation cost extinguishes the
merit of
heterogeneous networks in terms of cooperation.
In this situation, more connected players earn less
so that they are no longer spreaders of cooperation. Instead,
those with fewer contacts win and guide the
evolution. The participation cost, or the baseline payoff, is
irrelevant in homogeneous populations but is essential
for evolutionary games on heterogeneous networks.
\end{abstract}


\newpage

\section{Introduction}\label{sec:introduction}

A variety of creatures from
many small organisms to social animals including humans
show altruistic behavior. Altruism
occurs
even when being selfish is better for an individual, which
constitutes the social dilemma  \cite{Dawes80,Sugden86}.
Emergence of altruism under social
dilemmas can be explained by various mechanisms,
such as kin selection,
direct reciprocity,
indirect reciprocity, and group selection
\cite{Nowak06sci}.
In the Prisoner's Dilemma, altruism is also promoted by 
the viscosity of populations
\cite{Axelrod84,Nowak92}.
If players are aligned on a spatially
structured graph such as the square lattice, cooperators form
close-knit clusters of conspecifics to survive the invasion of selfish
defectors. Maintenance of such clusters is impossible in 
well-mixed populations modeled by the random
graph and the all-to-all connected network. 
Spatial structure
affects cooperation in other games as well
(e.g. \cite{Szabo02prl}). Particularly, spatial
structure can be detrimental to cooperation 
in the snowdrift game
\cite{Hauert02ijbc,Hauert04nat}.

To be more realistic, players often inhabit networks of social
contacts more complex than the square lattice, the random graph, and
the all-to-all connected network \cite{NewmanSIAM}.  First,
real social networks are small-world, implying the combination of
abundant localized interactions, as in the square lattice, and
sufficient shortcuts, as in the
random graph.  Second, players are heterogeneous in terms of the
number of contacts with others. An extreme case of this is the
scale-free network in which the number of neighbors is
distributed according to the power 
law \cite{Barabasi99}.
In contrast, the number of
neighbors of conventional networks
is the same for everybody (regular lattices and the
all-to-all connected network) or distributed with a narrow tail (so-called 
Erd\"{o}s-R\'{e}nyi
random graph). 
Even though not all social networks are scale-free,
the number of neighbors is naturally heterogeneous.

Recently, it was shown that heterogeneous networks promote
evolution of cooperation in 
the
Prisoner's Dilemma, the snowdrift game, and the stag hunt game.
Particularly, scale-free networks are strong amplifiers of altruism
\cite{Duran,Santos05prl,Santos06jeb,Santos06pnas,Santos06royal}.  

To explain the mechanism of enhanced cooperation,
let us introduce the notion of the temperature of players.  In
evolutionary graph theory, hot players are those replaced often by
others \cite{Lieberman}. By modifying this definition slightly, we
denote by hot players those who play often, namely, players with many
neighbors.  Cold players are those with a small number
of neighbors, such as leaves in a network. Hot players are allowed in
more rounds of the game than cold players per generation. If
the typical payoff obtained
by playing a game is positively biased, which 
is a common assumption adopted in the previous studies cited above
and others \cite{
Abramson,Ebel02,Ifti04,
Zimmermann05,Eguiluz05,Santos06plos} (but see
\cite{Pacheco06,Tomassini06} for other setups), it is largely worth
participation.  Then hot players earn more than cold players because
everybody earns positive `base' payoffs proportional to the number of
neighbors. As a result, hot players are more successful in
disseminating their strategies.  Particularly, cooperation once
employed by a hub is stable due to cooperative reactions in its
neighborhood. In addition, if hubs tend to be connected to each
other, as in the model proposed by Barab\'{a}si and Albert (BA model)
\cite{Barabasi99}, cooperators on hubs form loose clusters to take
advantage of a variant of spatial reciprocity
\cite{Santos05prl,Santos06pnas,Santos06royal}.

However, hot players are successful not owing to playing well but
owing to the connectivity.  In the present paper, we critically
re-examine the effect of heterogeneous networks on emergence of
cooperation.  In real lives, participation in the game may be
costly. A link to a neighbor implies building and maintaining
communication, and this cost has actually been modeled for studying
network formation \cite{Jackson,Bala,Goyal05}. Expensive
entry fees would dismiss the premium of hot players to change the
entire scenario.

We study two-person games on networks with participation costs.  For
simplicity, networks are assumed to be fixed in size and topology.
We show that there are
three regimes depending on how costly participation is.  First, when
participation is inexpensive,
cooperation is enhanced on heterogeneous
networks as discovered previously.
Second, when the participation cost is intermediate, the
effect of the local payoff structure, namely, the configuration of the
two-person game, and that of the network are comparable. Then altruism
does not develop.  Third, when participation is very costly, initial
strategies of cold players have long transients and 
propagate to hot players.
With small and
large participation costs, the network rather than the local payoff
structure determines evolutionary dynamics. In the intermediate
regime, evolution is most sensitive to the local payoffs.


\section{Model}

We compare effects of two types of networks on the evolution of
cooperation by means of Monte Carlo simulations.
A diluted well-mixed population is modeled by the
regular random graph in which each player has 8 neighbors that are
chosen randomly from the population. 
The heterogeneous networks are modeled by 
scale-free networks generated by the BA model,
in which the number of neighbors
denoted by $k$ follows the power law $p(k)\propto k^{-3}$
\cite{Barabasi99}.
The average number of neighbors in the
scale-free networks is set equal to 8 for fair comparison with the
regular random graph.  Both types of networks consist of $n=5000$
players.

To probe the network effect, we consider
only two simple strategies without memory, namely, unconditional
cooperation and unconditional defection.
The initial fraction of cooperators is set equal to
0.5.
In one generation, everybody participates in the two-person
game with all the neighbors. The payoff matrix will be specified in
the next section. 

Each player tends to copy successful strategies in its neighborhood.
We apply the update rule compatible with the replicator dynamics,
following the previous literature
\cite{Santos05prl,Santos06pnas,Santos06royal}.  Suppose that
player $x$ with $k_x$ neighbors has obtained generation-payoff
$P_x$. To update the strategy, $x$ selects a player $y$ among the
$k_x$ neighbors with equal probability ($=1/k_x$). Then $x$ copies
$y$'s strategy with probability $\left(P_y - P_x\right)$ $\big/$
$\left\{\max\left(k_x,k_y\right)\cdot \right.$
%
%
[uppermost payoff in one game - lowermost payoff in one game]
$\left.\right\}$ if $P_y > P_x$. The denominator is the normalization
constant so that the replacement probability ranges between 0 and 1.
If $P_y \le P_x$, the strategy of $x$ is unchanged. All the players
experience updating according to this rule synchronously. This
completes one generation.

Each evolutionary simulation consists of 5000 generations.  The final
fraction of cooperators denoted by $c_f$ is calculated as 
the average fraction of
cooperators based on the last 200 generations of 5 runs with different
initializations for each network and 5 different realizations of the
network. When the participation cost is not very large,
$c_f$ corresponds to values close to stochastic stationary values.
Otherwise, $c_f$ represents
transient values.

\section{Results}

\subsection{Prisoner's Dilemma}

We start with two well-known payoff matrices of the
simplified Prisoner's Dilemma. The first one is 
given by
\begin{equation}
\bordermatrix{
 & C & D \cr
C & 1 & 0 \cr
D & T & 0 \cr}. \;
\label{eq:pd1-basic}
\end{equation}
The entries of \EQ(\ref{eq:pd1-basic}) indicate the payoff that the row
player gains when playing against the column player.
The first (second) row and column
correspond to cooperation (defection).
The Prisoner's Dilemma arises when $T>1$, and
larger $T$ results in more defectors.
With participation cost $h$, the payoff matrix becomes
\begin{equation}
\left(\begin{array}{cc}
1-h & -h\\
T-h & -h
\end{array}
\right).
\label{eq:pd1-net}
\end{equation}
Note that introducing $h$ does not trespass the notion of the
Prisoner's Dilemma as far as $T>1$.

Actually, 
the game defined by \EQ(\ref{eq:pd1-basic})
or \EQ(\ref{eq:pd1-net}) lies on the boundary between
the snowdrift game analyzed in \SEC\ref{sub:sg} and
the Prisoner's
Dilemma. Therefore we
also examine another standard 
payoff matrix of the generic Prisoner's
Dilemma given by

\begin{equation}
\bordermatrix{
 & C & D \cr
C & b-c & -c \cr
D & b & 0 \cr}, \;
\label{eq:pd2-basic}
\end{equation}
where $b>c$. 
With the participation cost, \EQ(\ref{eq:pd2-basic}) is transformed to
\begin{equation}
\left(\begin{array}{cc}
1-r-h & -r-h\\
1-h & -h
\end{array}
\right),
\label{eq:pd2-net}
\end{equation}
where $r=c/b$.
In the following, 
we refer to numerical results
for the payoff matrix given by \EQ(\ref{eq:pd1-net}).
The results that are 
qualitatively the same as the following are obtained for 
\EQ(\ref{eq:pd2-net}),
as shown in Supplementary figure~\ref{fig:supp1}(b-d).

The 
fraction of cooperators $c_f$ is not affected by $h$ on the regular
random graph (Supplementary figure~\ref{fig:supp1}(a)). 
Because each player has the
same number of neighbors, participation cost does not differentiate
the players.  In contrast, $h$ drastically affects
$c_f$ for the scale-free networks,
as shown in \FIG\ref{fig:pd}(a). 
In \FIG\ref{fig:pd}(b),
$c_f$ for the scale-free networks relative to
$c_f$ for the regular random graph is plotted.
We identify three
qualitatively different regimes in terms of $h$, which
roughly correspond to (I) $h< 0.24$, (II) $0.24\le h < 2$, and (III)
$h\ge 2$. The transition between (II) and (III) is fairly gradual.

\subsubsection{Regime (I): Costless Participation}

When $h< 0.24$, participation is inexpensive, and
hot players such as hubs are strong
competitors regardless of the strategies of their cold neighbors.
The payoff of a player increases linearly with the number of
contacts, as shown by thin lines with positive slopes in
\FIG\ref{fig:pd_flip}(a).
In particular, cooperation spreads from hot cooperators to their cold
neighbors, the local density of cooperators increases, and
hot cooperators gain more by mutual cooperation.
Cooperation triggered by hot players is self-promotive.
Defective hot players may also win for
a moment.  However, defection then prevails in their neighborhood so
that hot defectors can no longer exploit the neighbors because of mutual
defection. This results in a null
generation-payoff of hot defectors so that they can be outperformed by
their cold neighbors. A hot player sticks to cooperation but
not to defection.

In sum, heterogeneous networks enhance altruism, which
recovers the previous work corresponding to $h=0$
\cite{Santos05prl,Santos06jeb,Santos06pnas,Santos06royal}.  Note that
this regime extends to $h<0$, that is, when gifts are given for
participation so that everyone wins a positive payoff.

To illuminate on different dynamics of hot and cold players, we
measure how often players flip the strategy. 
The average number of flips throughout the evolutionary run
(including the contribution from both transients and stationary states)
is shown in \FIG\ref{fig:pd_flip}(b).
Colder players experience more flips when $h <
0.24$ (thin lines).  They myopically follow what hotter players do both 
in transients and in
stationary states. Cooperation on hubs is
stabilized in an early stage, yielding
less flips of hotter players.

\subsubsection{Regime (II): Moderately Expensive Participation}

Interestingly, cold players spread their strategies to hot players
when $h\ge 0.24$ (regimes (II) and (III)), which is opposite to what
happens in regime (I). As a result, enhanced cooperation
diminishes, even with a relatively small participation cost.

Regime (II) is defined by small to intermediate $h$ ($0.24\le h<
2$).  Now the local payoff structure as well as the network topology
is relevant.  When $h=0.3$, scale-free networks surpass the regular
random graph in terms of the number of cooperators only for $1 \le
T\le 1.4$. When $h=0.6$, this range shrinks to $1\le T\le 1.1$. The
privilege of scale-free networks is entirely lost when $h=1$.  In
regime (II), the payoff of a player linearly
decreases with the number of
contacts (thick lines with negative slopes in \FIG\ref{fig:pd_flip}(a)).
Consistent with this, 
hot players flip strategies more often than cold players
(thick lines in \FIG\ref{fig:pd_flip}(b)).
Hubs no longer conduct the dynamics.

\subsubsection{Regime (III): Costly Participation}

\begin{figure}
\begin{center}
\includegraphics[height=2in,width=2in]{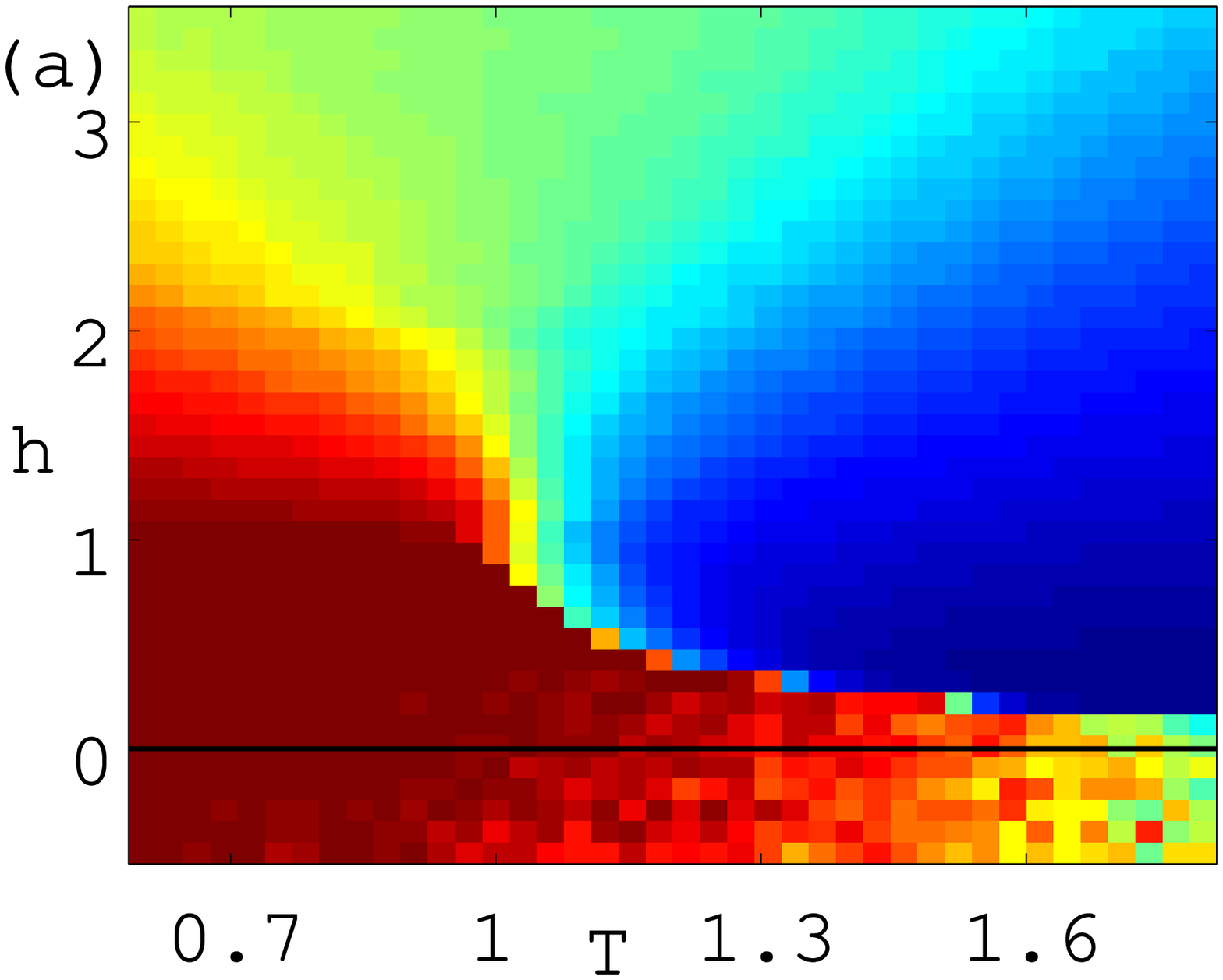}
\includegraphics[height=2in,width=0.4in]{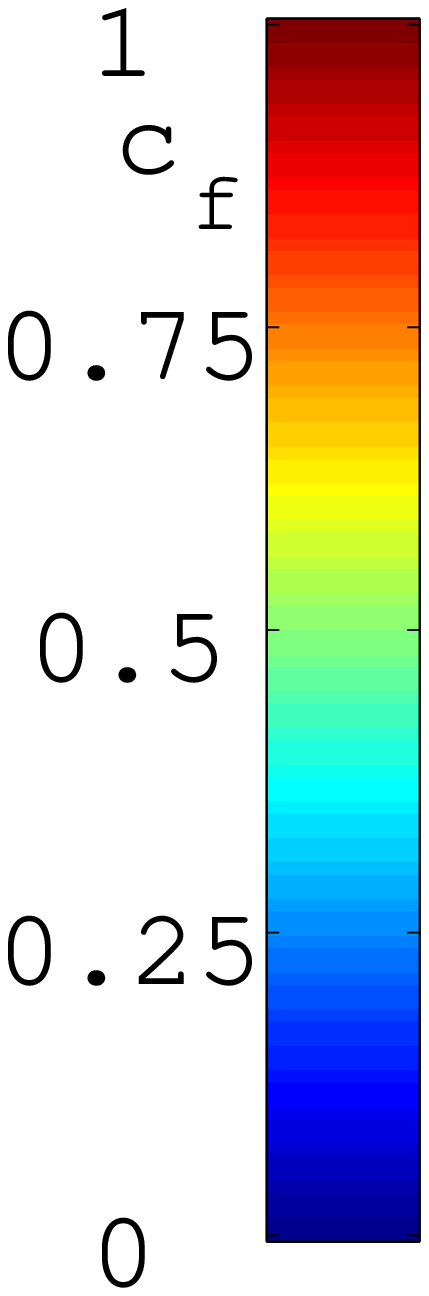}
\includegraphics[height=2in,width=2in]{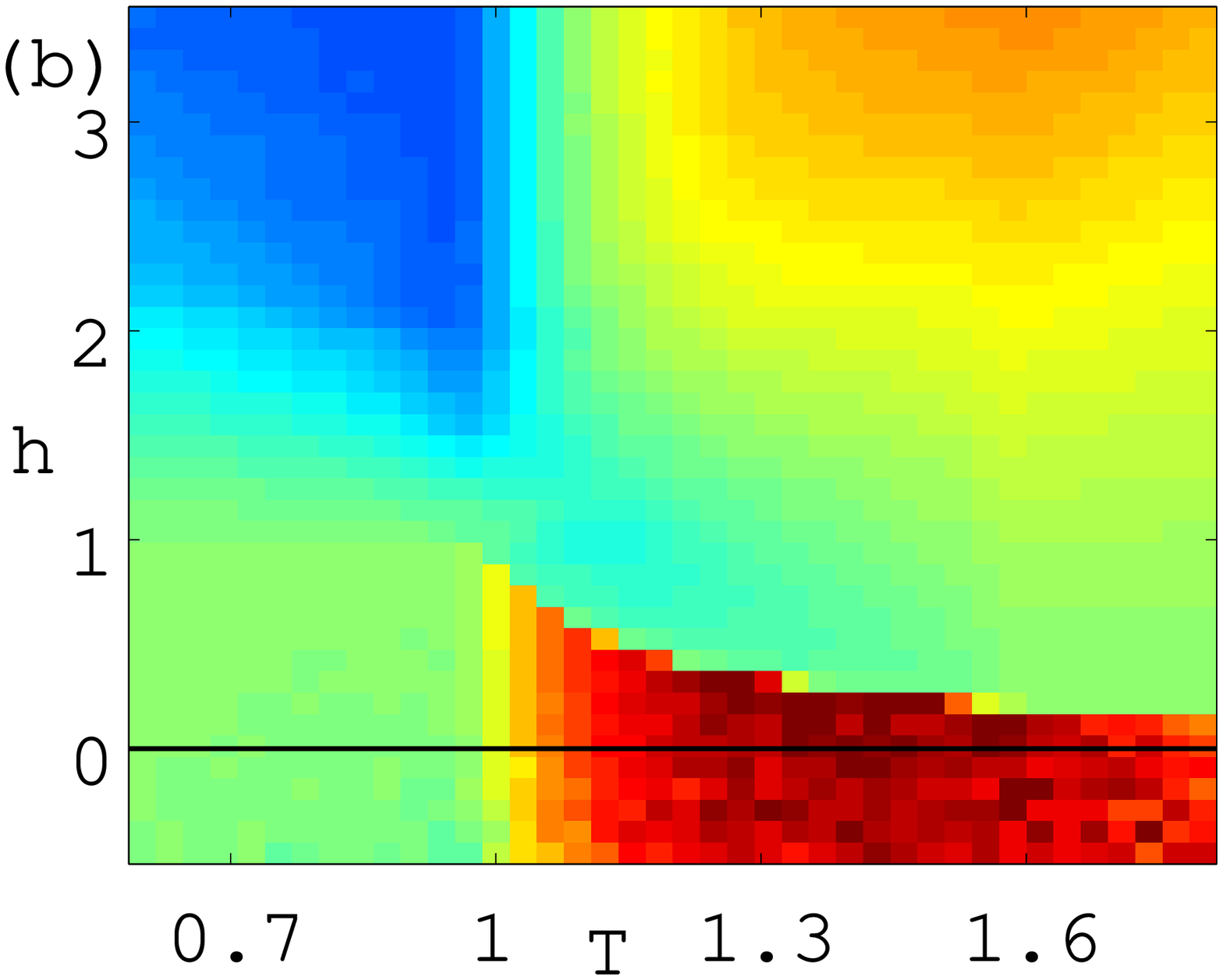}
\includegraphics[height=2in,width=0.4in]{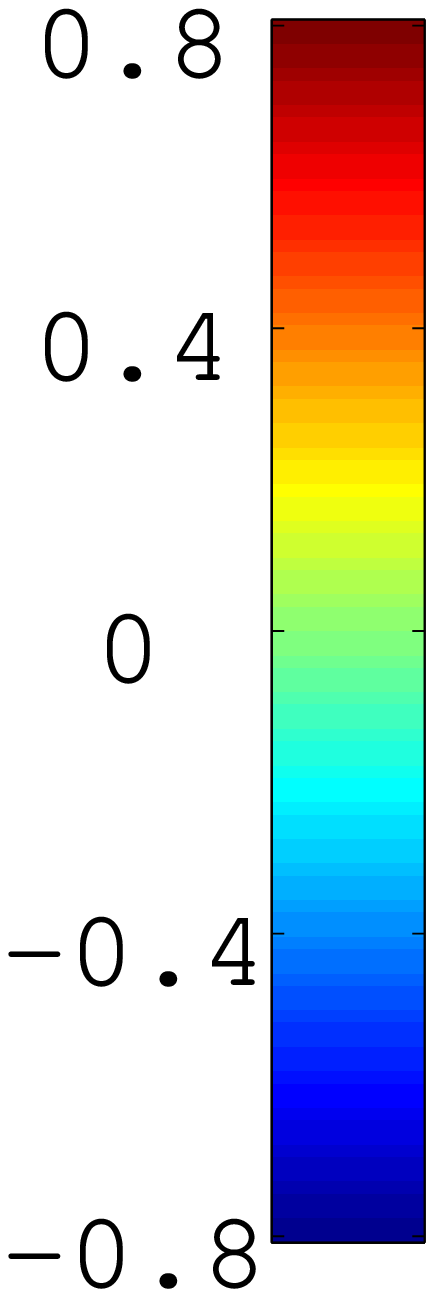}
\caption{}
\label{fig:pd}
\end{center}
\end{figure}

\begin{figure}
\begin{center}
\includegraphics[height=2.5in,width=4in]{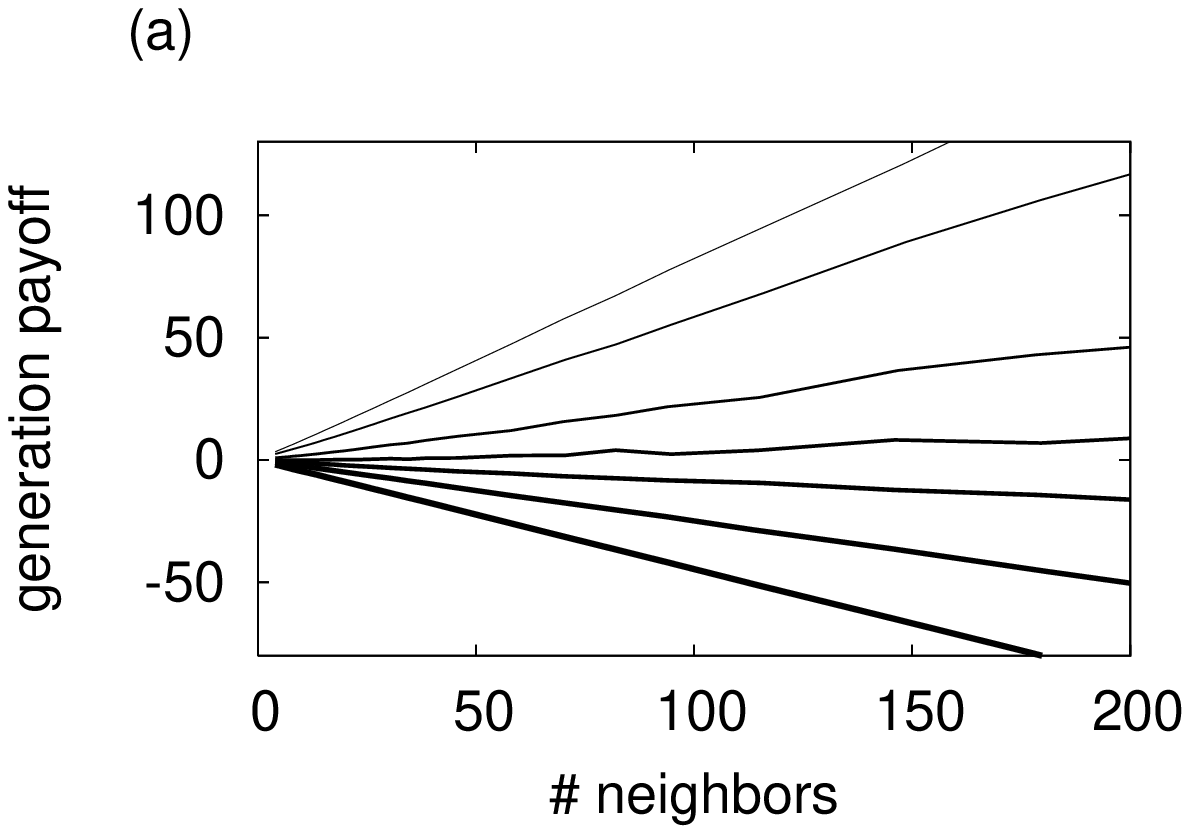}
\includegraphics[height=2.5in,width=4in]{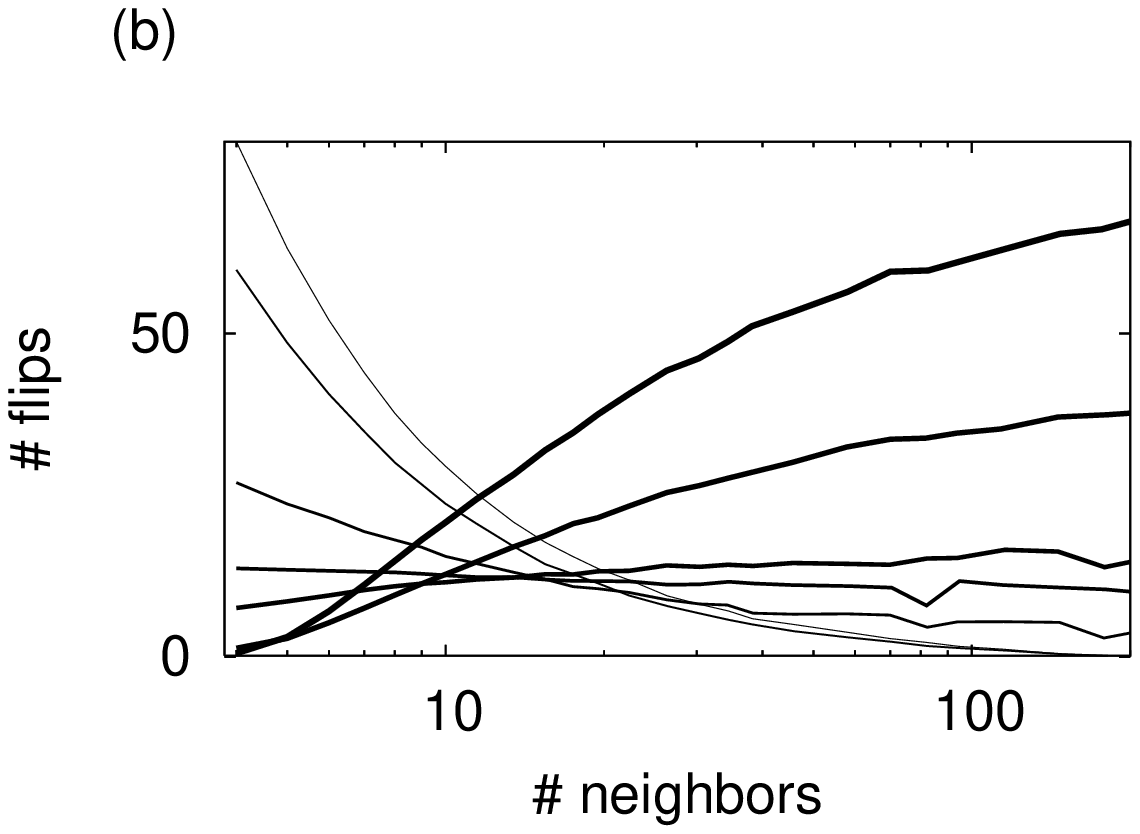}
\caption{}
\label{fig:pd_flip}
\end{center}
\end{figure}

When $h$ is roughly greater than 2, participation is really
costly. Then cold players with any strategy surpass hot players and
govern the dynamics.  This is not because cold players are tactical
but because they play less often and lose less than hot players do.
Strategies of cold players are not often replaced owing to the small
number of neighbors. Consequently, cold players persist in their
initial strategies, actually up to $>300000$ generations. In the
meantime, strategies of cold players tend to spread to hot
neighbors. As a result, $c_f$, which is measured at 5000 generations,
is almost the same as the initial fraction of cooperators ($=0.5$)
regardless of $T$ (\FIG\ref{fig:pd}(a)). Figure~\ref{fig:pd}(b)
indicates that $c_f$ is larger for the scale-free networks than for
the random graph, but this is due to long transients. After a very
long period, the fraction of cooperators will be smaller than $c_f$
shown in \FIG\ref{fig:pd}(a) to eradicate the spurious advantage of
scale-free networks.


\subsection{Snowdrift Game}\label{sub:sg}

The snowdrift game, which is also known as the chicken game or the
hawk-dove game \cite{Sugden86,Hauert04nat}, is illustrated
as a situation of two drivers caught in a snowdrift.  For the two cars
to get out, which is equivalent to payoff $\beta$ $(more than 1)$ for each
driver, the snow must be shoveled away.  A total effort of unity must
be invested to this task.  Two players may cooperate to share the cost
so that each pays $1/2$, or one player may cover the full
cost. Otherwise, both players miss
$\beta$.  Cooperation benefits not only the opponent but also the
focal player itself.  Different from the Prisoner's Dilemma in
which defection is beneficial, in the snowdrift game,
cooperation can deserve even when the opponent
defects.

If the participation is free, the payoff matrix of the 
snowdrift game 
is given by
\begin{equation}
\bordermatrix{
 & C & D \cr
C & \beta-1/2 & \beta-1 \cr
D & \beta & 0 \cr}. \;
\label{eq:sg-basic}
\end{equation}
In this case, heterogeneous networks reinforce evolution of cooperation
as in the Prisoner's Dilemma
\cite{Santos05prl,Santos06pnas,Santos06royal}.
Without 
dismissing the structure of the snowdrift game,
the participation cost translates the payoff matrix to the following form:
\begin{equation}
\left(\begin{array}{cc}
\beta-1/2-h & \beta-1-h\\
\beta-h & -h
\end{array}
\right).
\label{eq:sg-net}
\end{equation}

As shown in Supplementary figure~\ref{fig:supp2}, the participation cost $h$ does not
influence $c_f$ on the regular random graph.
The fraction of cooperators converges
to a value close to
the theoretical estimate $c_f=1-r$, where $r = 1/(2\beta-1)$
is the cost-to-benefit ratio \cite{Sugden86,Hofbauerbook,Hauert04nat}.
If cooperation is
relatively costly with a small $\beta$ (large $r$), cooperators
decrease in number.

\begin{figure}
\begin{center}
\includegraphics[height=2in,width=2in]{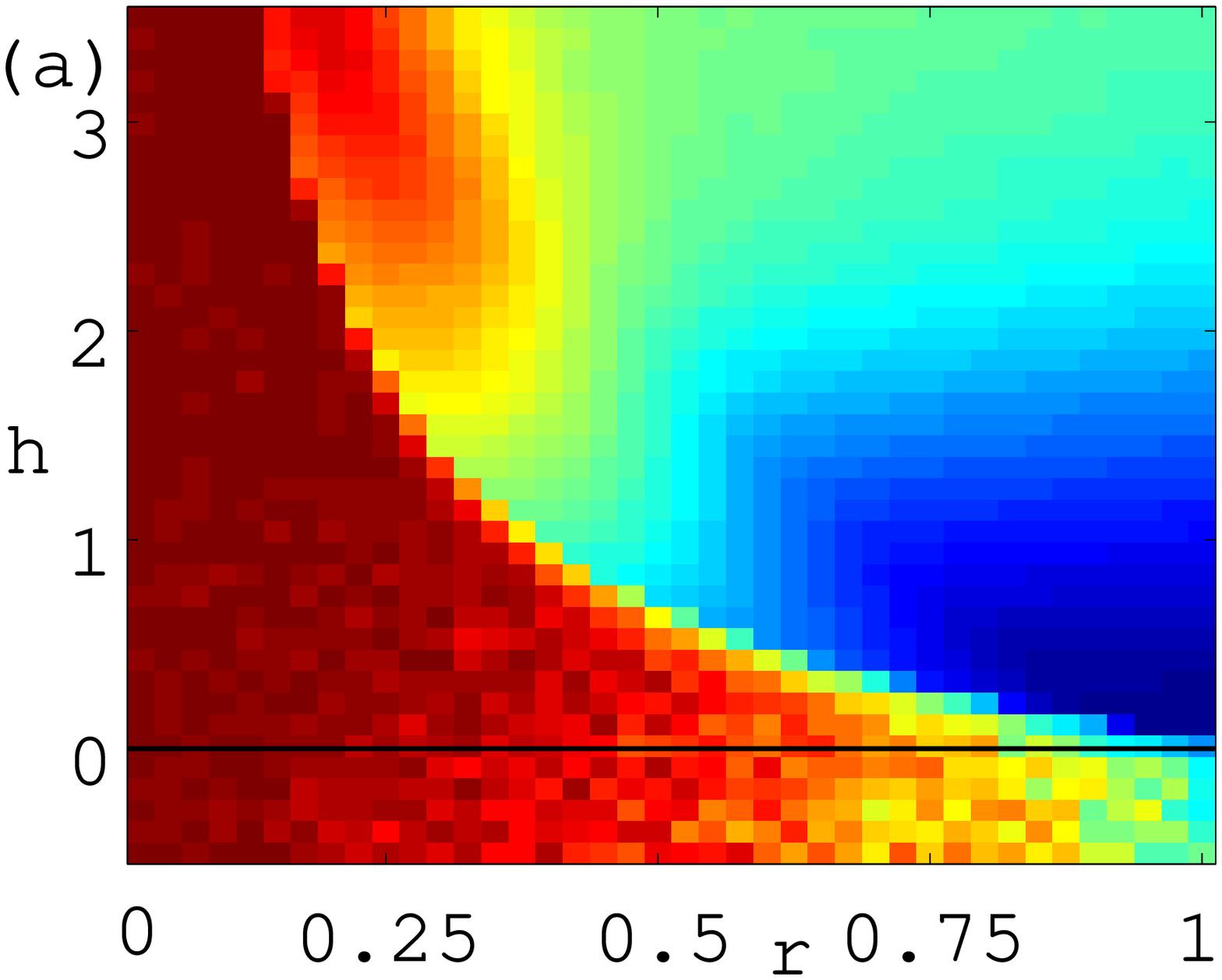}
\includegraphics[height=2in,width=2in]{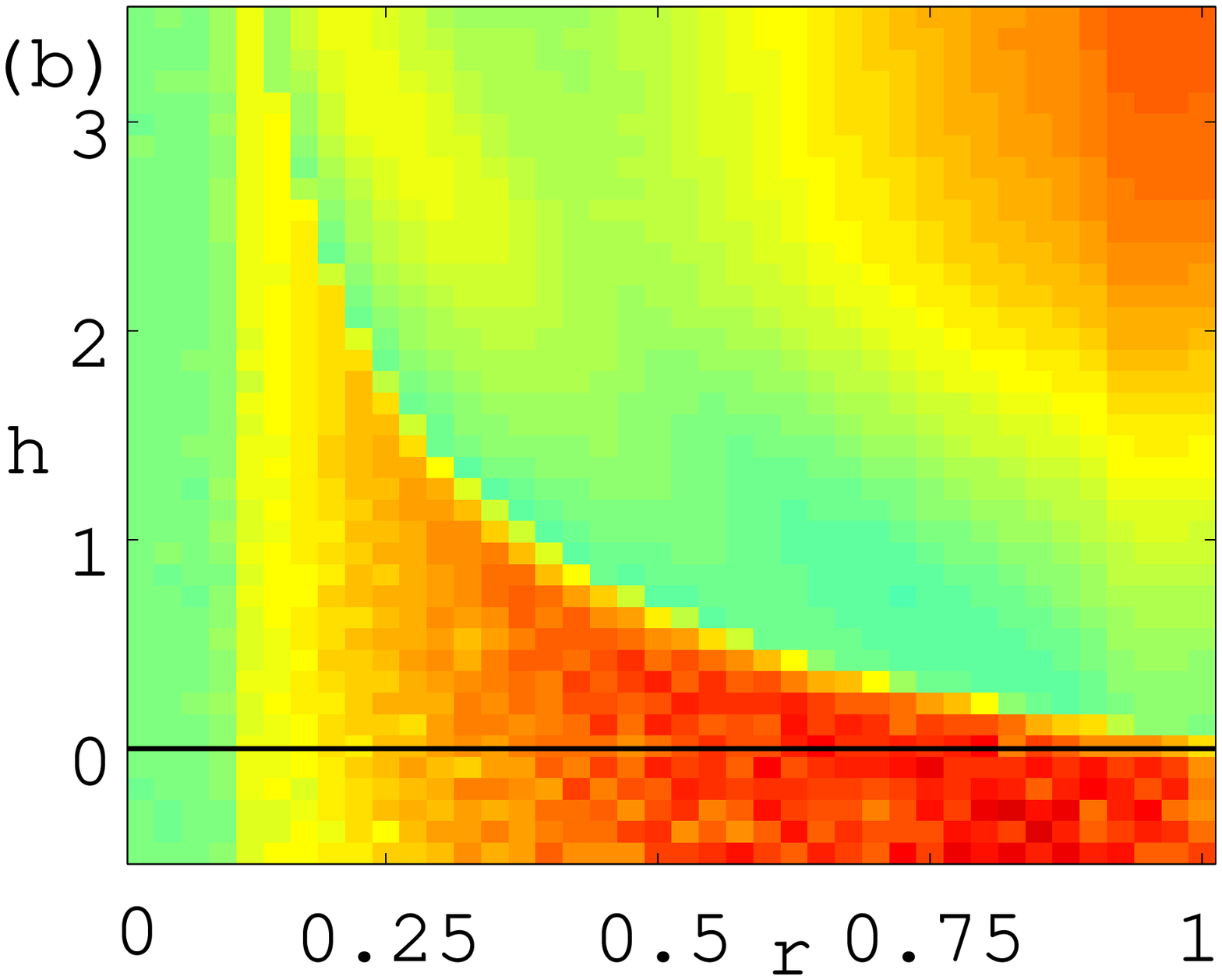}
\caption{}
\label{fig:sg}
\end{center}
\end{figure}

On heterogeneous networks, 
how the fraction of cooperators varies as a function of
$r$ depends on the participation cost.
We again find three regimes as shown in \FIG\ref{fig:sg}.
The scale-free networks host more cooperators than
the regular random graph
only when $h$ is near zero or
negative (regime (I)).
The advantage of the scale-free networks is neutralized
by intermediate $h$ (roughly speaking, $h\cong 1$), which defines
regime (II).
Note that the reduction of cooperation is
not as much as that for the Prisoner's Dilemma.  With large $h$
(roughly speaking, $h\ge 2$),
$c_f$ is rather insensitive to the local payoff structure
due to long transients (regime (III)).

\subsection{General Two-person Games}\label{sub:general}

With the participation cost incorporated,
general symmetrical two-person games with two strategies
are represented by
\begin{equation}
\bordermatrix{
 & C & D \cr
C & R-h & S-h \cr
D & T-h & P-h \cr}. \;
\label{eq:full}
\end{equation}
In accordance with the previous sections,
we denote by cooperation (defection)
the strategy corresponding to the first (second) row
and column.
%
%
As $T$ increases, players are tempted to
defect, and $c_f$ decreases.
As $S$ decreases, players would refuse cooperation to avoid
exploitation by defectors. Therefore, $c_f$ decreases.
The Prisoner's Dilemma, the snowdrift game, and the stag hunt game,
are defined by $T>R>P>S$,
$T>R>S>P$, and $R>T>P>S$, respectively.
The Prisoner's Dilemma usually accompanies
another condition $2R>T+S$ so that
mutual cooperation is more beneficial 
than alternating unilateral cooperation.

Multiplying each element of \EQ(\ref{eq:full})
by a common constant
modifies just the time scale of evolution.
Accordingly, there are three free
parameters in the payoff matrix, which 
are chosen to be $T$, $S$, and $h$, while we set
$R=1$ and $P=0$.

In \FIG\ref{fig:T_and_S}(a), $c_f$ for $h=0$ is plotted in the $T$-$S$
parameter space for the
regular random graph.  As expected, the number of cooperators
decreases with $T$ and increases with $S$. The results are
independent of $h$.
For the scale-free networks, we plot $c_f$ in
\FIG\ref{fig:T_and_S}(b-e) for four values of $h$ (also see
Supplementary figure~\ref{fig:supp3} for the direct comparison of $c_f$ for the
scale-free networks and $c_f$ for the regular random graph). We
confirm the existence of the three regimes, extending the results
shown in \FIGS\ref{fig:pd} and \ref{fig:sg}.  First, as shown in
\FIG\ref{fig:T_and_S}(b), scale-free networks promote evolution of
cooperation when the participation is costless \cite{Santos06pnas}.
Cooperation is strengthened in the Prisoner's Dilemma ($T>1$, $S<0$),
the snowdrift game ($T>1$, $S>0$), and also the stag hunt game ($T<1$,
$S<0$).  Second, the advantage of the heterogeneity is lost for a wide
range of $T$ and $S$ when $h=0.5$ (\FIG\ref{fig:T_and_S}(c)) and $h=1$
(\FIG\ref{fig:T_and_S}(d)). Third, the transient is very long
when participation is costly ($h=2$). This allows defectors to survive
for a long time
even without dilemma ($S>0, T<1$) and considerable cooperators to
survive under the Prisoner's Dilemma
(\FIG\ref{fig:T_and_S}(e)).

\section{Discussion}

We have discovered that the participation cost, which is irrelevant in
well-mixed populations and on homogeneous networks including the
regular lattices, casts a dramatic effect on evolutionary dynamics on
heterogeneous networks. When
participation is nearly free (regime (I)), heterogeneous networks
promote cooperation
\cite{Duran,Santos05prl,Santos06jeb,Santos06pnas,Santos06royal}.  This
is because the cooperation on hot players
is robust against invasion of whatever strategies of cold
players. Even if a cold player is good at exploiting cooperators
in the neighborhood, the aggregated payoff would be much smaller than
that of a hot player that would earn a lot just by playing the game many
times.  Hot players are leaders, and cold players are myopic
followers.  However, this phenomenon is not robust against variation
in participation costs, which is consistent with the loss of
cooperation under positive affine transformation of the payoff matrix
\cite{Tomassini06}.  When the participation cost is intermediate
(regime (II)), cooperators do not really increase and even decrease on
heterogeneous networks.  When participation is costly (regime (III)),
hot players myopically 
follow cold players.
In regimes (I) and (III), not local payoff structure
but network structure governs evolution. The local payoffs are
relevant only in regime (II), for which cooperation is not enhanced by
heterogeneous networks.  Regimes (II) and (III), which have been
largely unexplored, may be relevant in, for example, environmental
problems, political conflicts, and human relationships.  In these
situations, players are often forced to play and incur participation
costs.  In other words, the best one could get may
be the least disastrous, but not really wonderful, solution.

For replicator-type and many other update rules
in well-mixed populations,
evolution is invariant under
uniform addition of a constant to the payoff matrix
(e.g. participation cost).
Then a game in regimes (II) and (III)
can be translated into a game in regime (I).
However, this operation is disallowed for heterogeneous
networks. Because
multiplying the payoff matrix
by a positive constant
alters just the evolutionary time scale in either case,
there are two free parameters for two-person games in homogeneous
populations, whereas there are three of them
in heterogeneous populations
(e.g. $S$, $T$, and $h$ as used in
\FIG\ref{fig:T_and_S}). 

The present results can be generalized in some aspects.  First,
two-person games can be asymmetrical. Second, the update rule does not
have to be of a replicator type \cite{Hofbauerbook,Ohtsuki06nat} as
far as the reproduction rate is not extremely nonlinear in
the generation payoff. Third, 
large noise would blur but would not break down the
three regimes.
For example, irrational actions may cause
hot cooperators, which are stable with small participation costs and
small noise, to defect and elicit upsurges of defectors
nearby. However, noise does not overturn the fact that hot (cold)
players are better off for small (large) participation
costs.   Fourth, network models can be arbitrary.
For example, 
the scale-free networks based on the configuration model (without
growth and preferential attachment) and the Erd\"{o}s-R\'{e}nyi random
graph promote altruism in regime (I), albeit to a lesser extent than
the BA model \cite{Santos05prl,Santos06pnas,Santos06royal}. 
These heterogeneous networks as well as the BA model
do not enhance altruism in regimes (II)
and (III) (data not shown).

The present results also have limitations. 
First, we have only assumed memoryless strategies, namely,
unconditional cooperators and unconditional defectors.  Second,
the networks have been static.  In coevolutionary dynamics in which
players form and sever links as well as play games, only regime (I)
has been considered, a main conclusion being enhanced cooperation
\cite{Skyrms00,
Zimmermann05,Eguiluz05,Santos06plos}.  In
regime (III), for example,
everybody must sever links to be loners \cite{Goyal05}
(for the role of loners, also refer to
\cite{Hauert02sci}).
%
%
Third, we have used the additive payoff
\cite{Nowak94ijbc,Abramson,Ebel02,Ifti04,Santos05prl,Santos06jeb,Duran,Santos06pnas,Santos06royal}. An
alternative is to use the average payoff, or division of the
generation payoff of each player by the number of neighbors
\cite{Kim02pre,
Santos06jeb,Taylor06
}.
%
%
The average payoff is not affected by the number of neighbors and
hampers the enhanced altruism on heterogeneous networks
\cite{Santos06jeb,Tomassini06}. We have adopted the additive payoff
scheme because its intuitive meaning is clearer than the average payoff.
These topics warrant for future work.



\section*{Acknowledgments}
We thank Hisashi Ohtsuki and Eizo Akiyama
for their valuable discussions and critical reading
of the manuscript.

\newpage

Figure captions

\bigskip

Figure 1: The Prisoner's Dilemma on the scale-free networks with
participation costs. The payoff matrix is represented by
\EQ(\ref{eq:pd1-net}). 
(a) The final fraction of cooperators $c_f$. (b) $c_f$
for the scale-free networks shown in (a) minus $c_f$ for the regular
random graph shown in Supplementary figure~\ref{fig:supp1}a.

\bigskip

Figure 2: The effect of the number of neighbors in the
Prisoner's Dilemma on the scale-free networks.  (a) The average
generation payoff and (b) the average number of flips are
plotted in terms of the number of neighbors that a player has.
Note that the log scale is
used for the abscissa of (b) for clarity. The lines
correspond to $h=0$ (thinnest line), 0.2, 0.23, 0.24, 0.25, 0.3, and
0.5 (thickest line).  The payoff matrix is given by
\EQ(\ref{eq:pd1-net}) with $T=1.5$.

\bigskip


Figure 3: The snowdrift game on the scale-free networks with
participation costs.  (a) $c_f$, and (b) $c_f$ for the scale-free
networks shown in (a) minus $c_f$ for the regular random graph shown
in Supplementary figure~\ref{fig:supp2}.

\bigskip

Figure 4: $c_f$ in the $T$-$S$ space for (a) the regular random graph
and the scale-free networks with (b) $h=0$, (c) $h=0.5$, (d) $h=1$,
and (e) $h=2$.

\newpage

Supplementary figure 1: (a) The final fraction of cooperators $c_f$ in
the Prisoner's Dilemma on the regular random graph.  The payoff matrix
is represented by \EQ(\ref{eq:pd1-net}).
The results for the payoff matrix \EQ(\ref{eq:pd2-net}) 
are shown in 
(b-d). (b) $c_f$ for the regular random graph, (c) $c_f$ for the
scale-free networks, and (d) $c_f$ for the scale-free networks minus
$c_f$ for the regular random graph.

\bigskip

Supplementary figure 2: $c_f$ for the
snowdrift game on the regular random graph.

\bigskip

\begin{figure}
\begin{center}
\includegraphics[height=2in,width=2in]{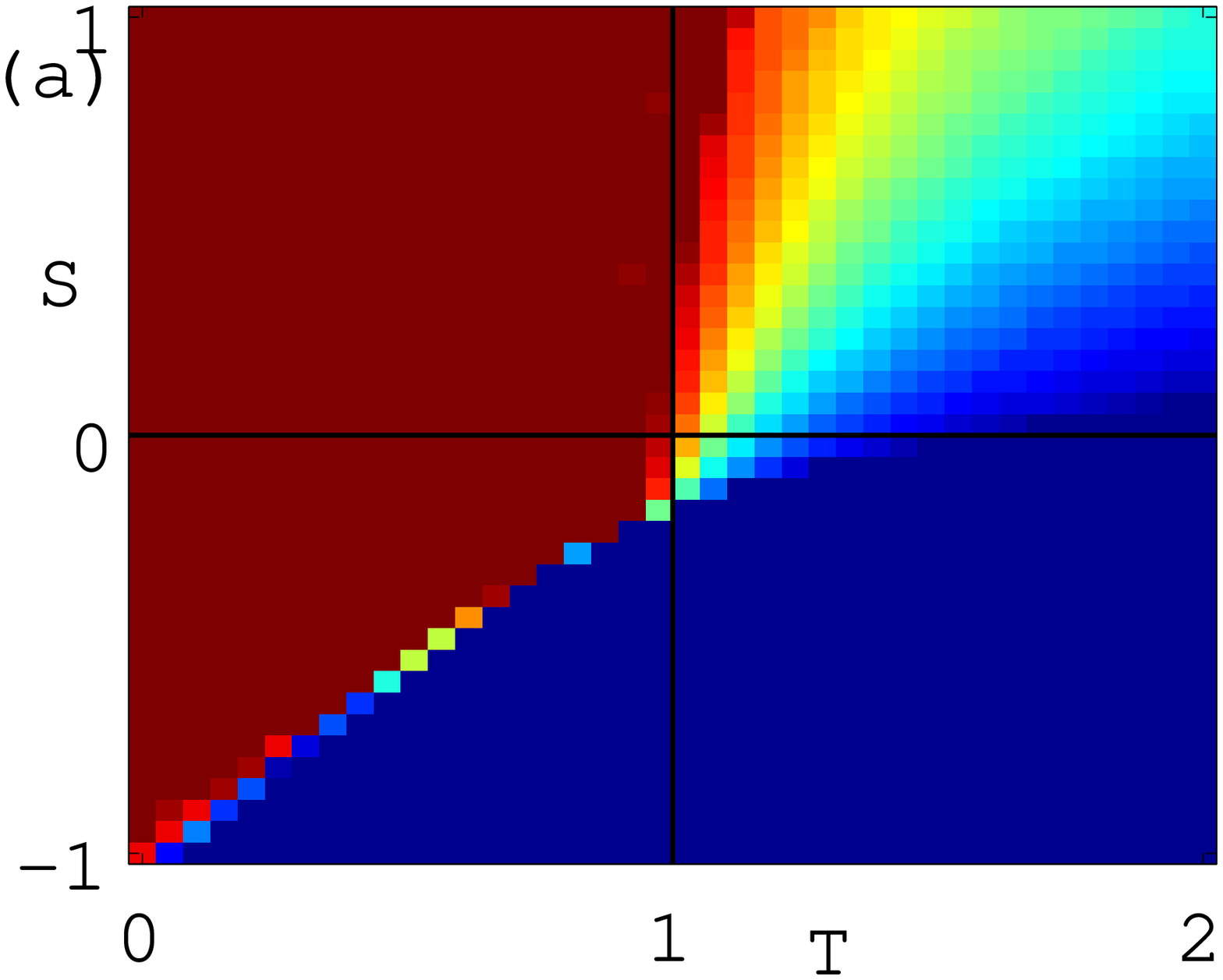}
\includegraphics[height=2in,width=2in]{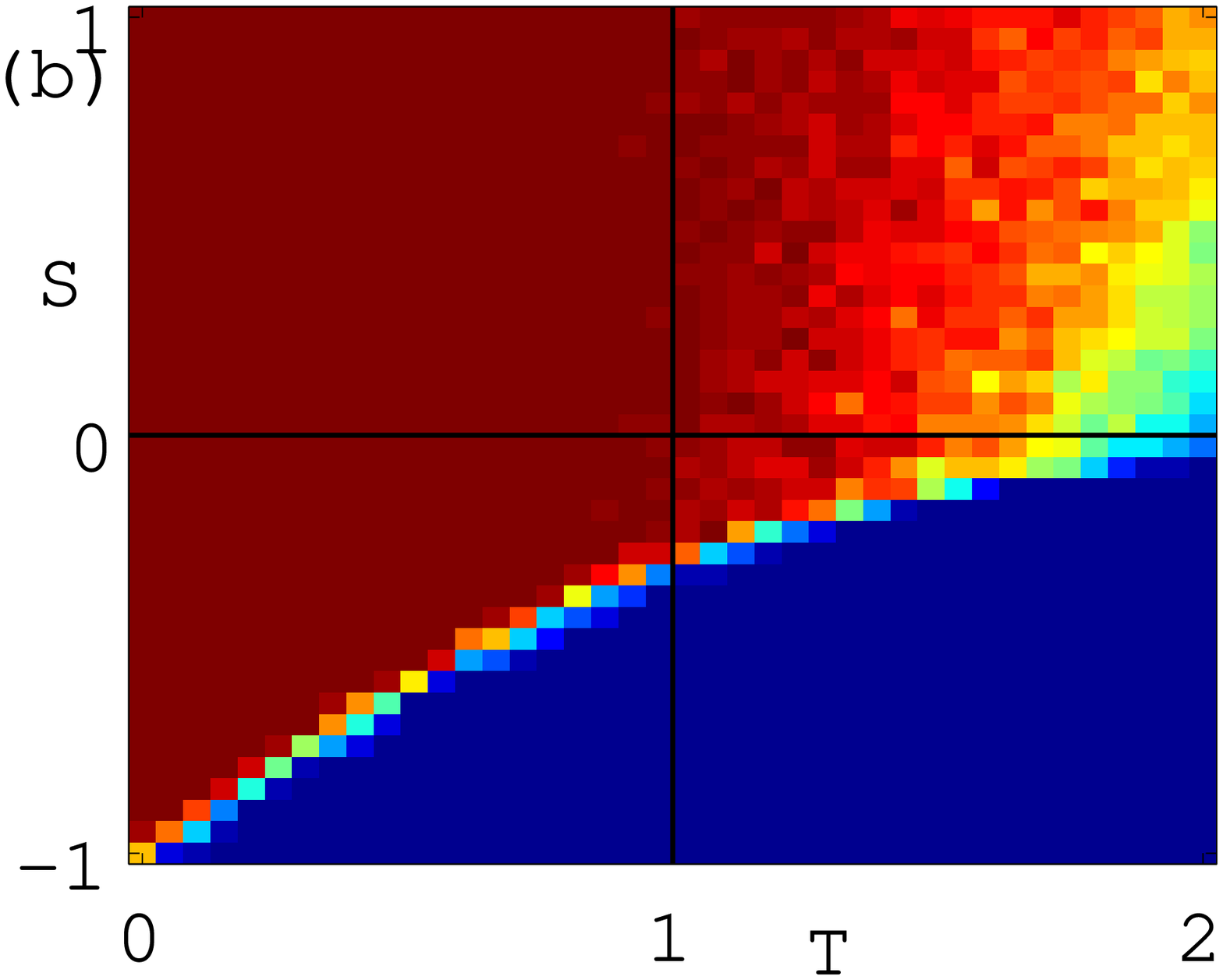}
\includegraphics[height=2in,width=2in]{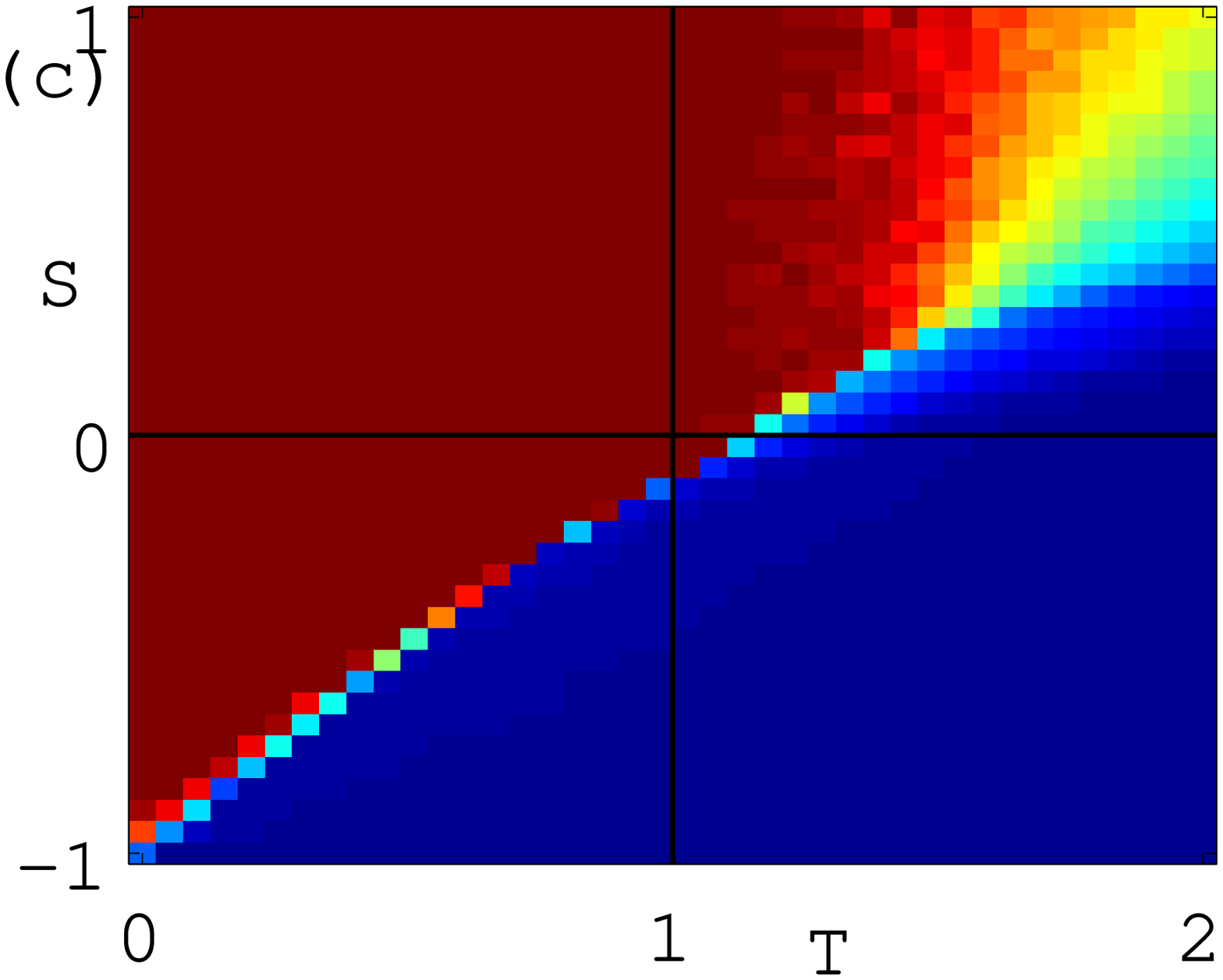}
\includegraphics[height=2in,width=2in]{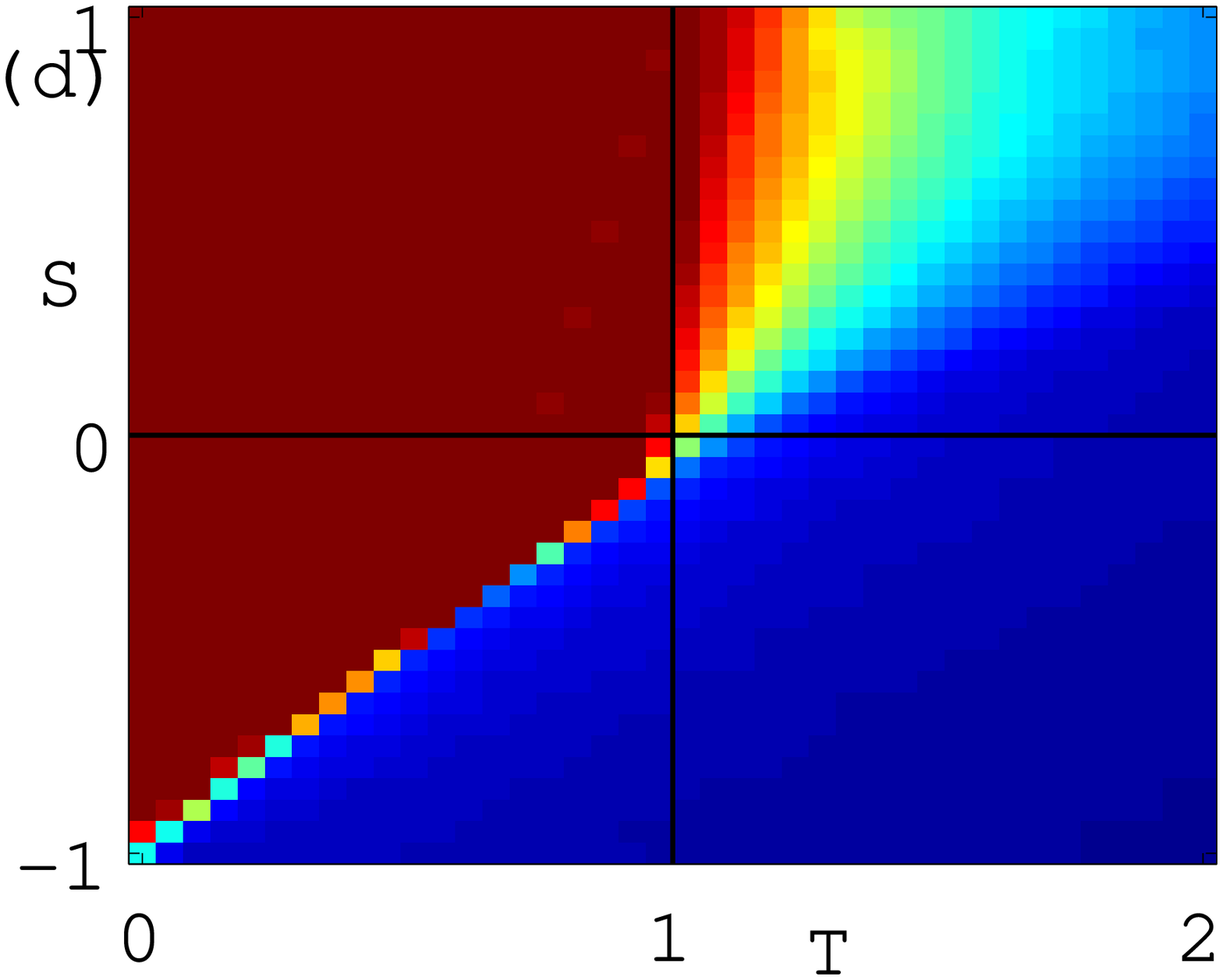}
\includegraphics[height=2in,width=2in]{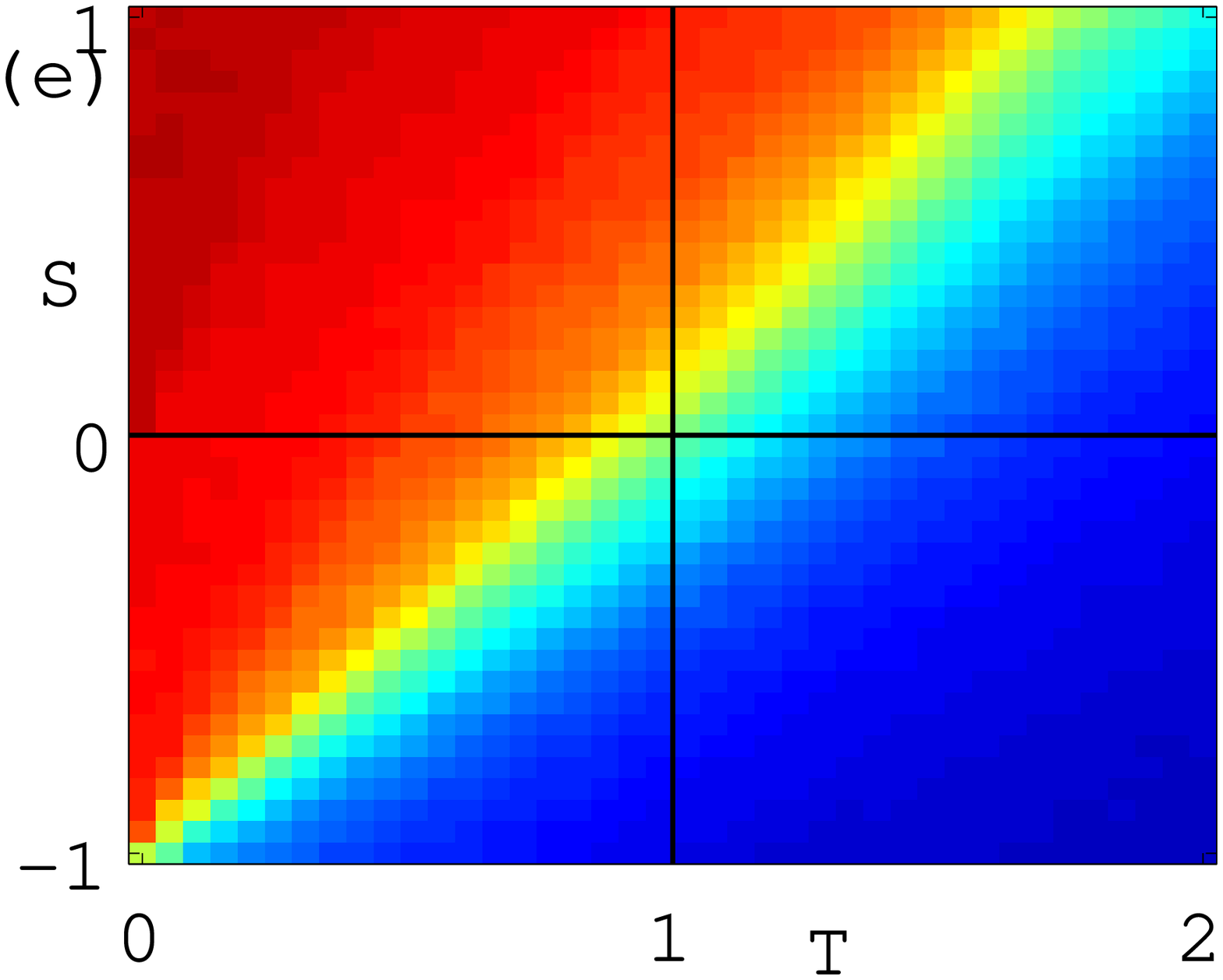}
\caption{}
\label{fig:T_and_S}
\end{center}
\end{figure}

Supplementary figure 3:
$c_f$ for the
scale-free networks (\FIG\ref{fig:T_and_S}(b-e)) minus
$c_f$ for the regular random graph (\FIG\ref{fig:T_and_S}(a)) in
the $T$-$S$ space.
(a) $h=0$, (b) $h=0.5$, (c) $h=1$, and (d) $h=2$.

\newpage

\clearpage

\clearpage

\clearpage

\clearpage

\renewcommand{\figurename}{Supplementary figure}
\setcounter{figure}{0}

\begin{figure}
\begin{center}
\includegraphics[height=2in,width=2in]{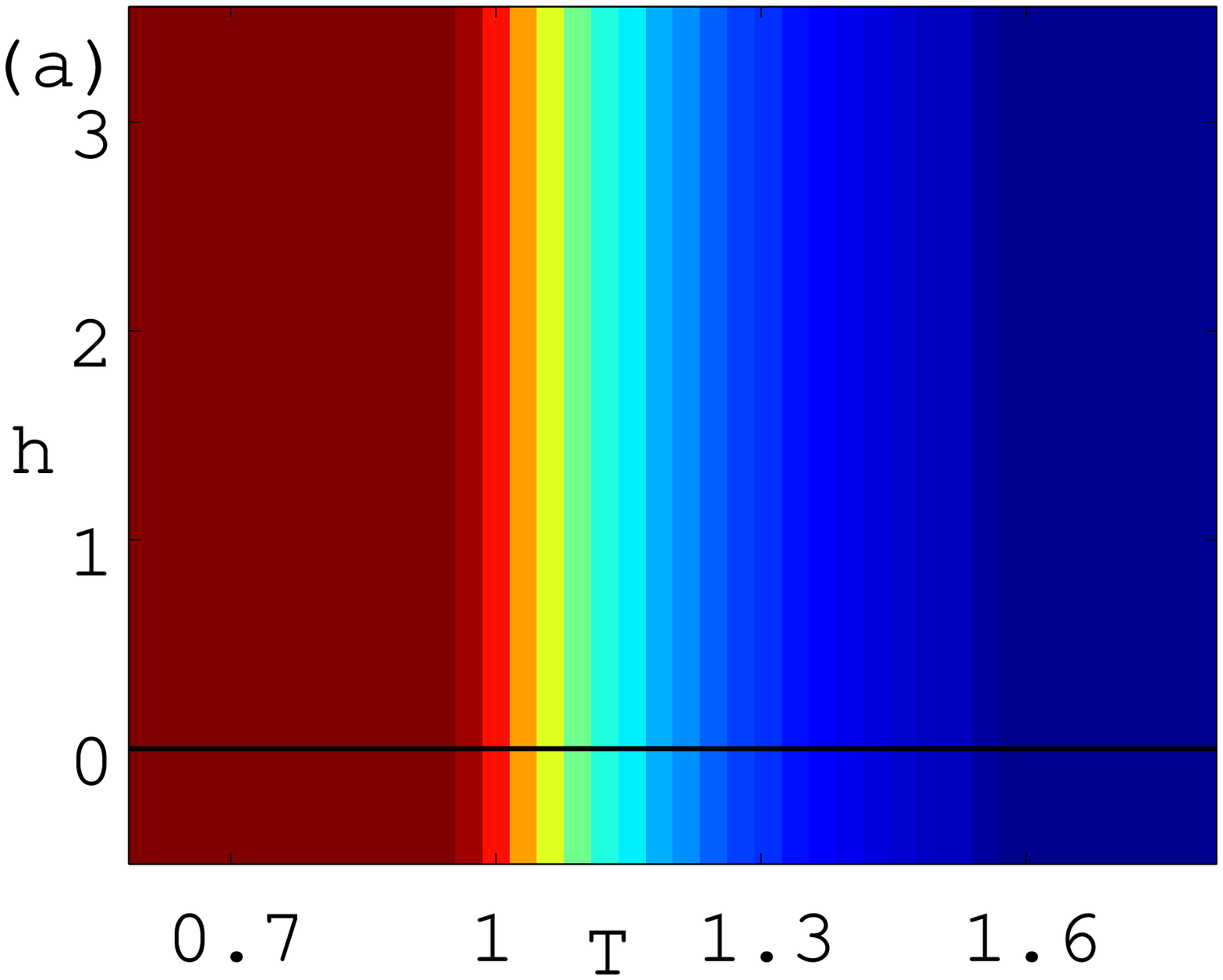}
\includegraphics[height=2in,width=0.4in]{cf_gauge.eps}
\includegraphics[height=2in,width=2in]{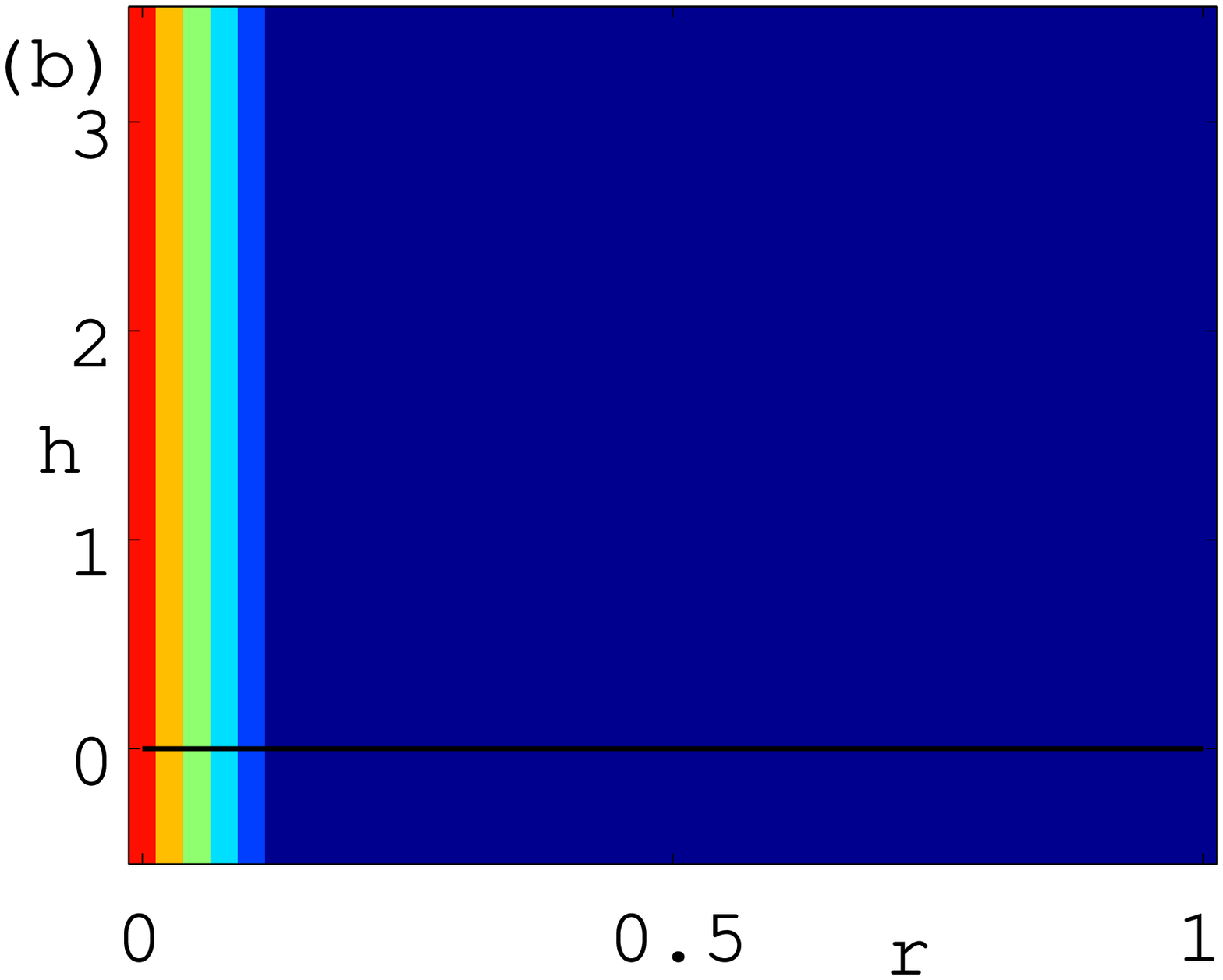}
\includegraphics[height=2in,width=2in]{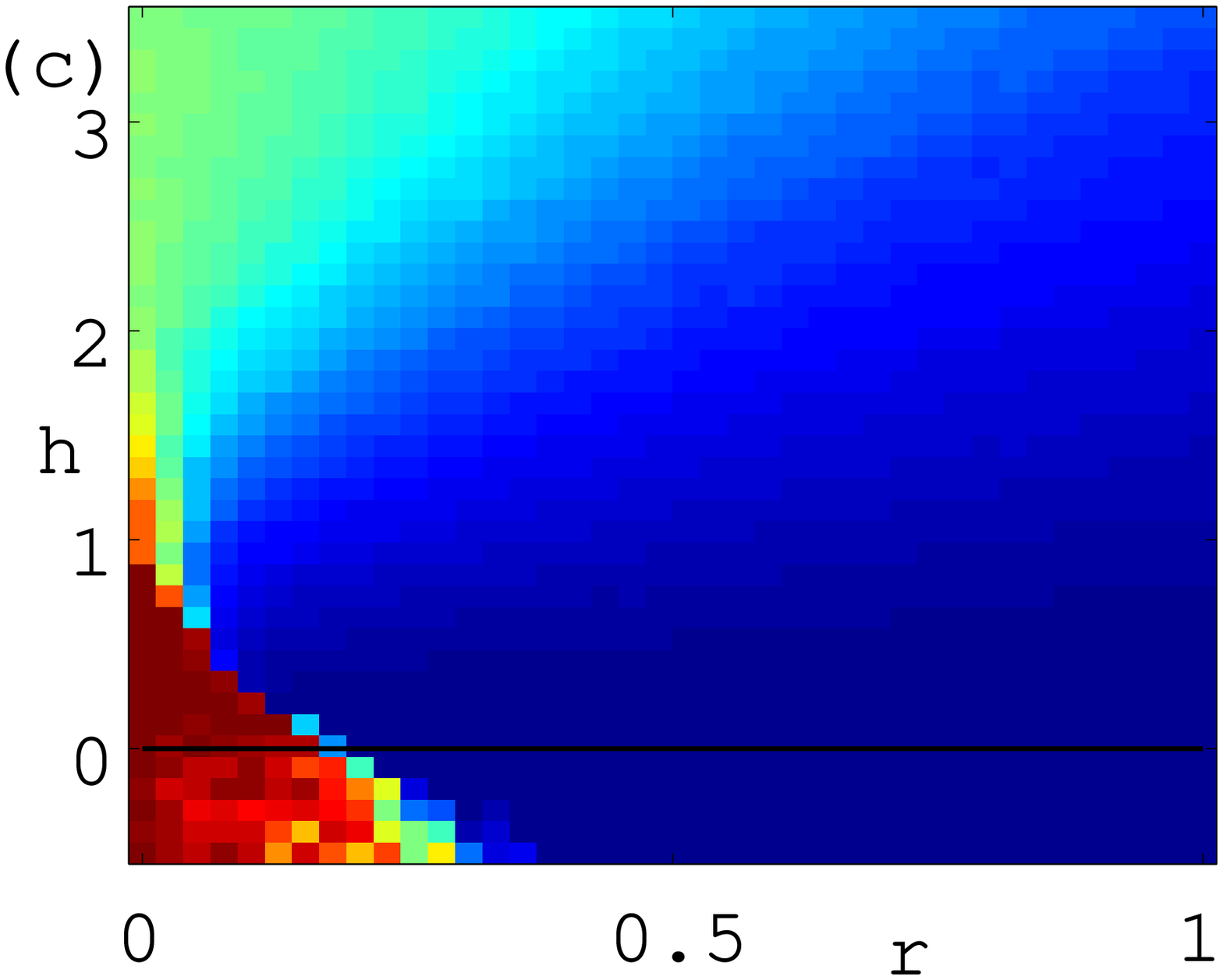}
\includegraphics[height=2in,width=2in]{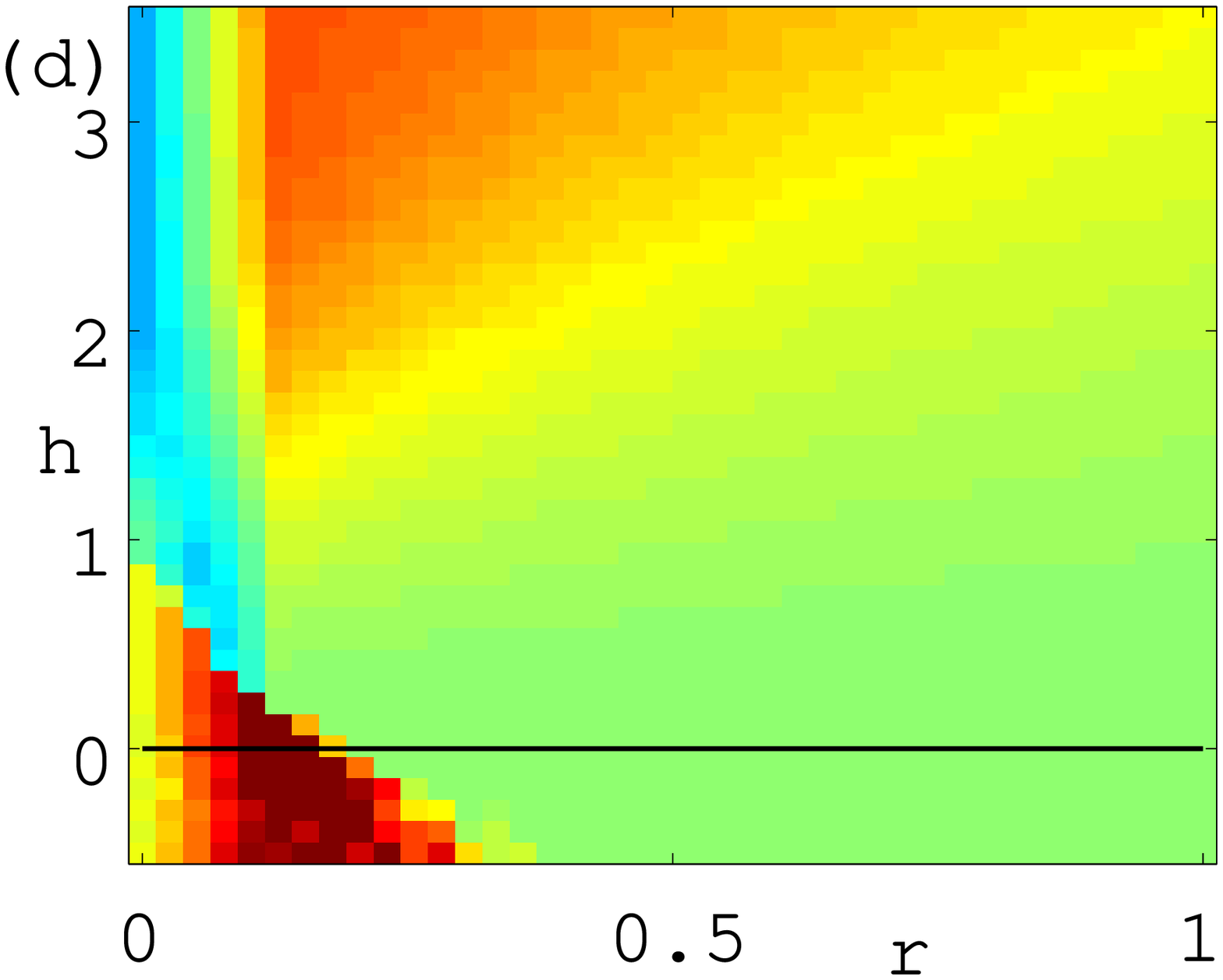}
\includegraphics[height=2in,width=0.4in]{diff_gauge.eps}
\caption{}
\label{fig:supp1}
\end{center}
\end{figure}

\clearpage

\begin{figure}
\begin{center}
\includegraphics[height=2in,width=2in]{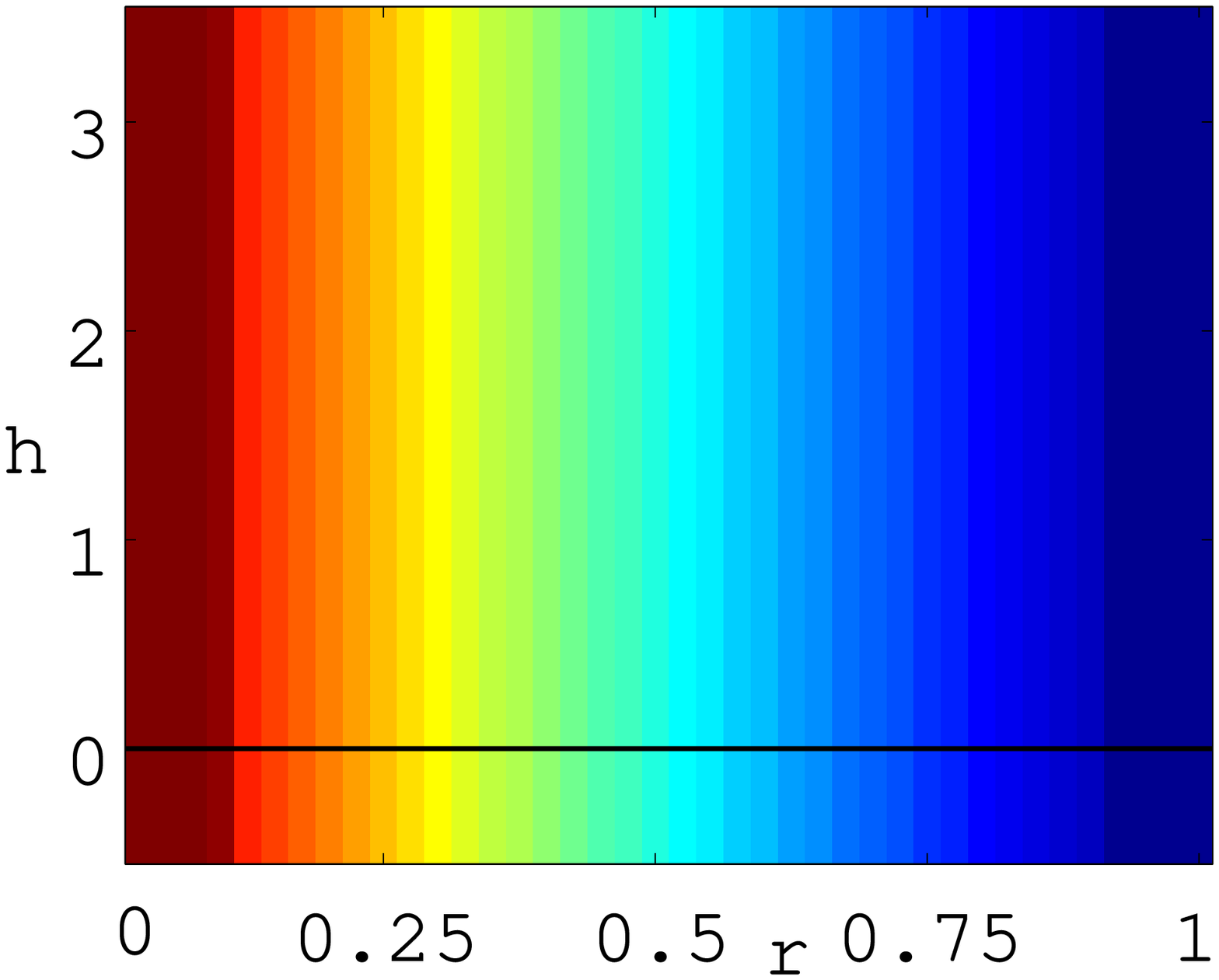}
\caption{}
\label{fig:supp2}
\end{center}
\end{figure}

\clearpage

\begin{figure}
\begin{center}
\includegraphics[height=2in,width=2in]{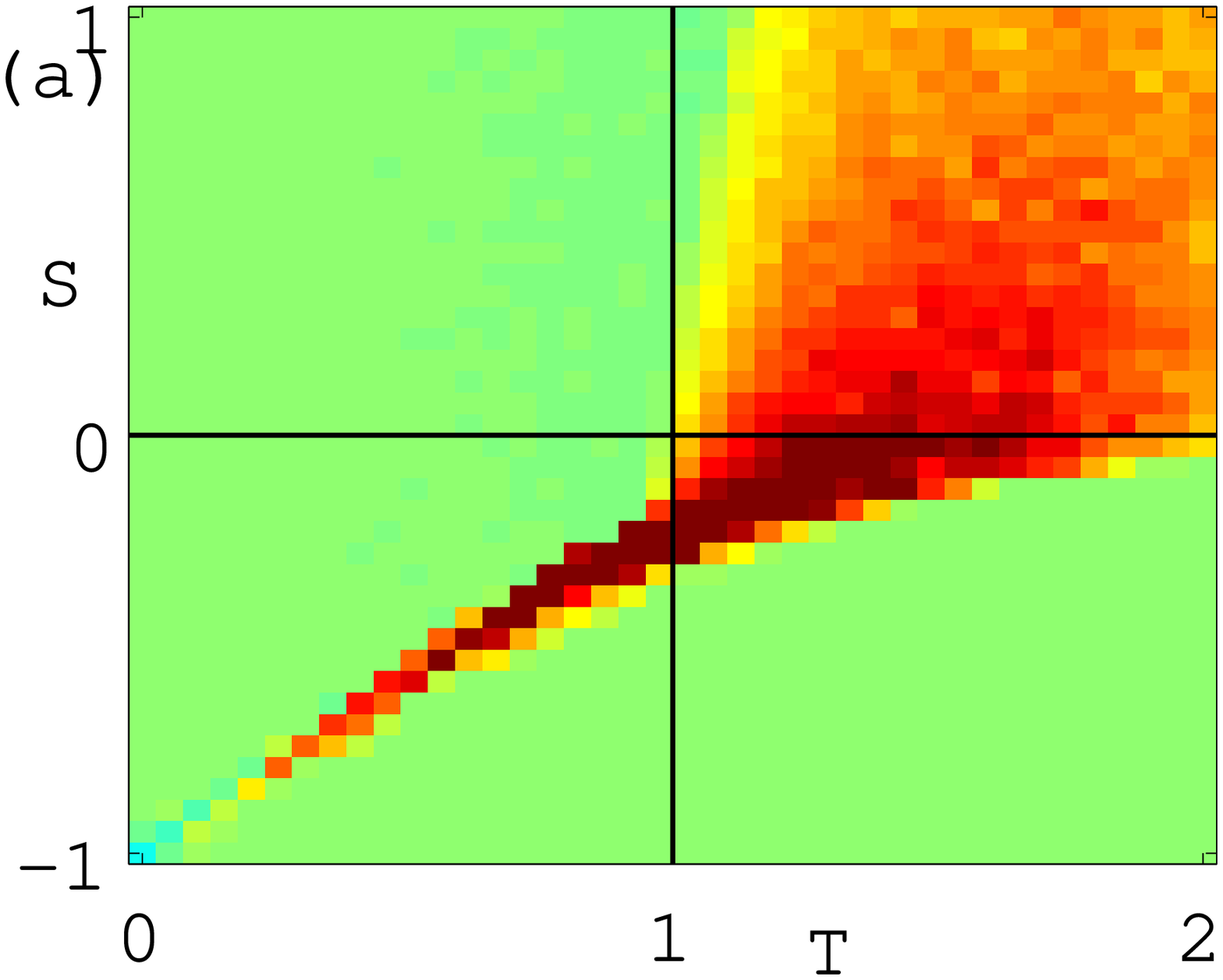}
\includegraphics[height=2in,width=2in]{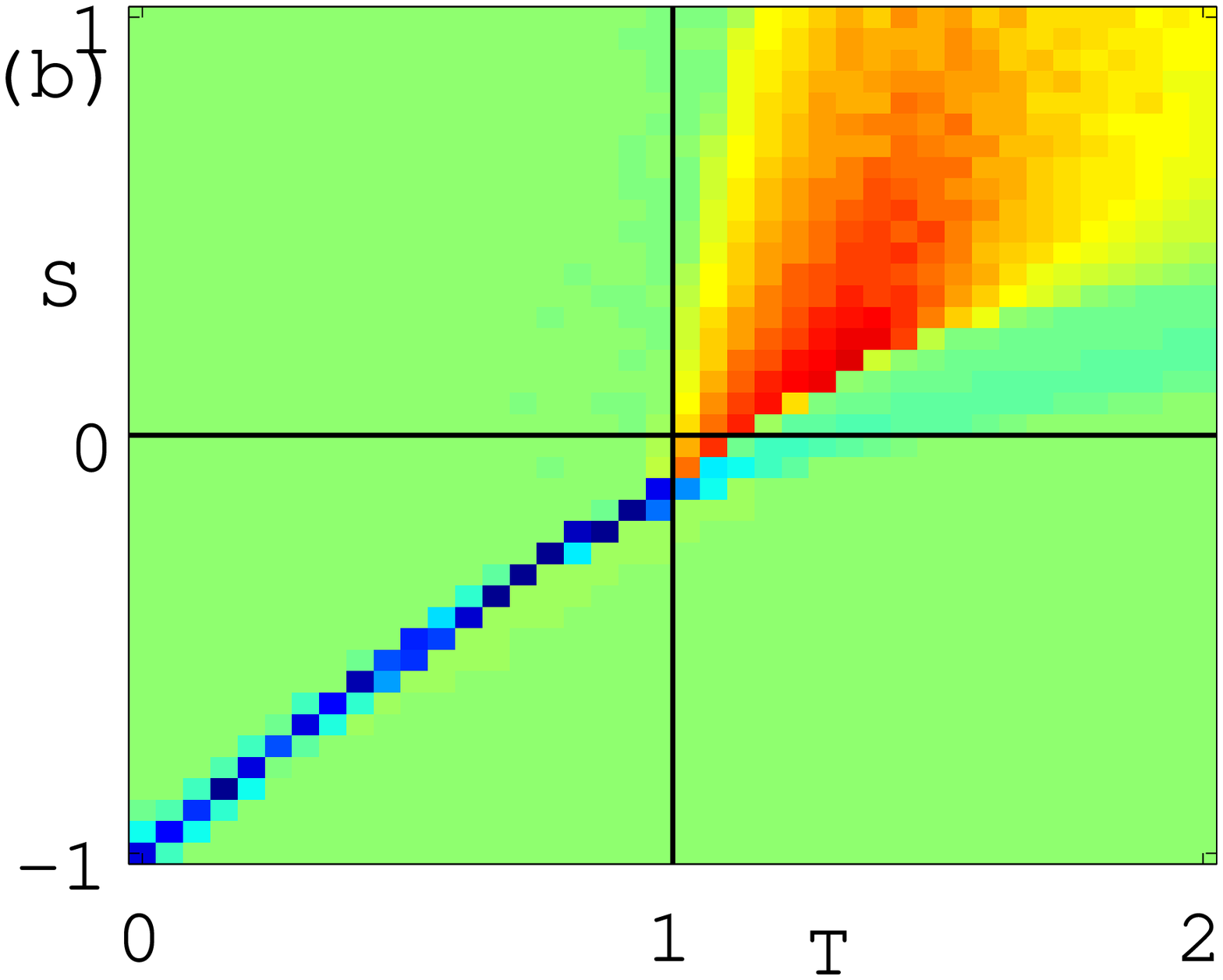}
\includegraphics[height=2in,width=2in]{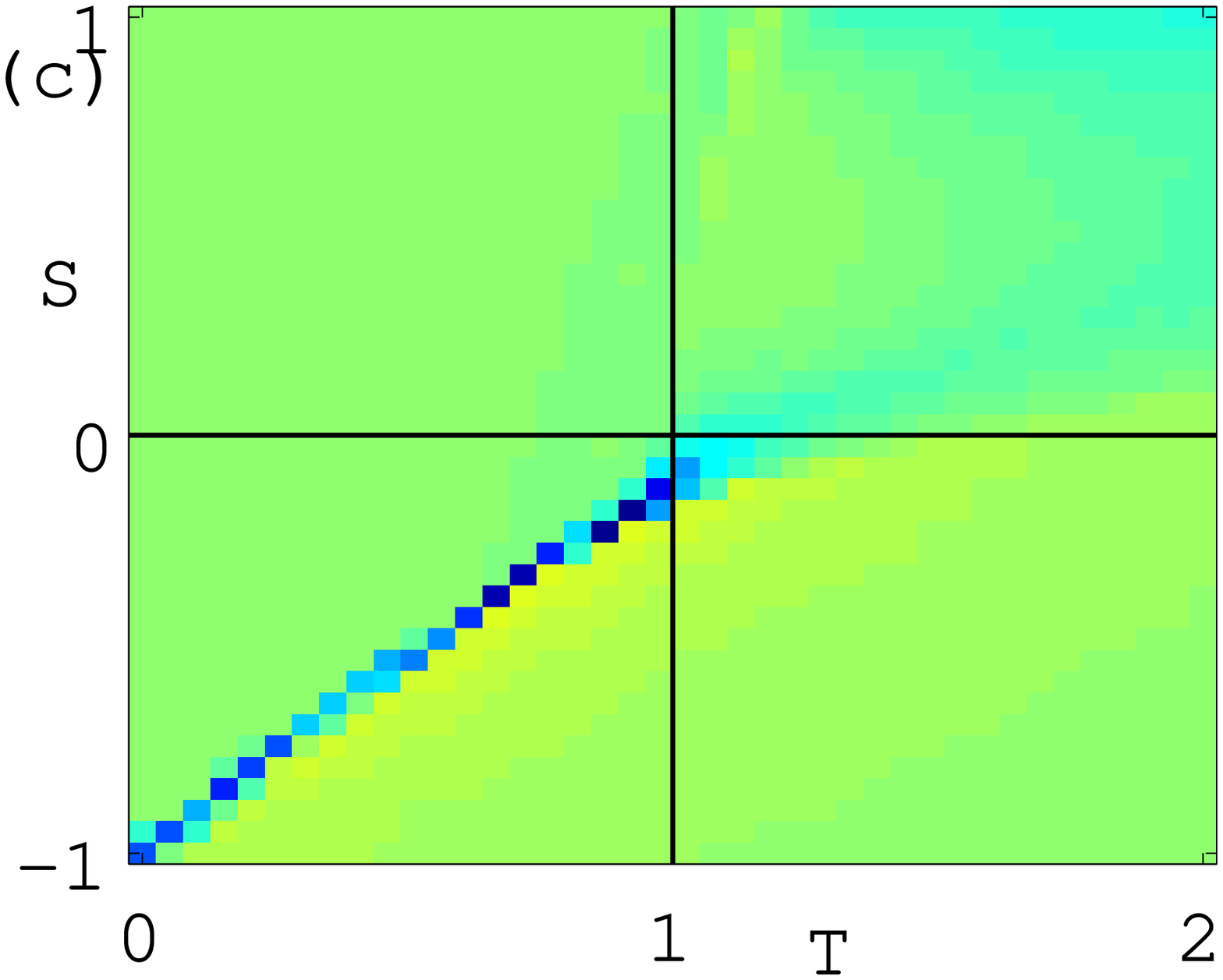}
\includegraphics[height=2in,width=2in]{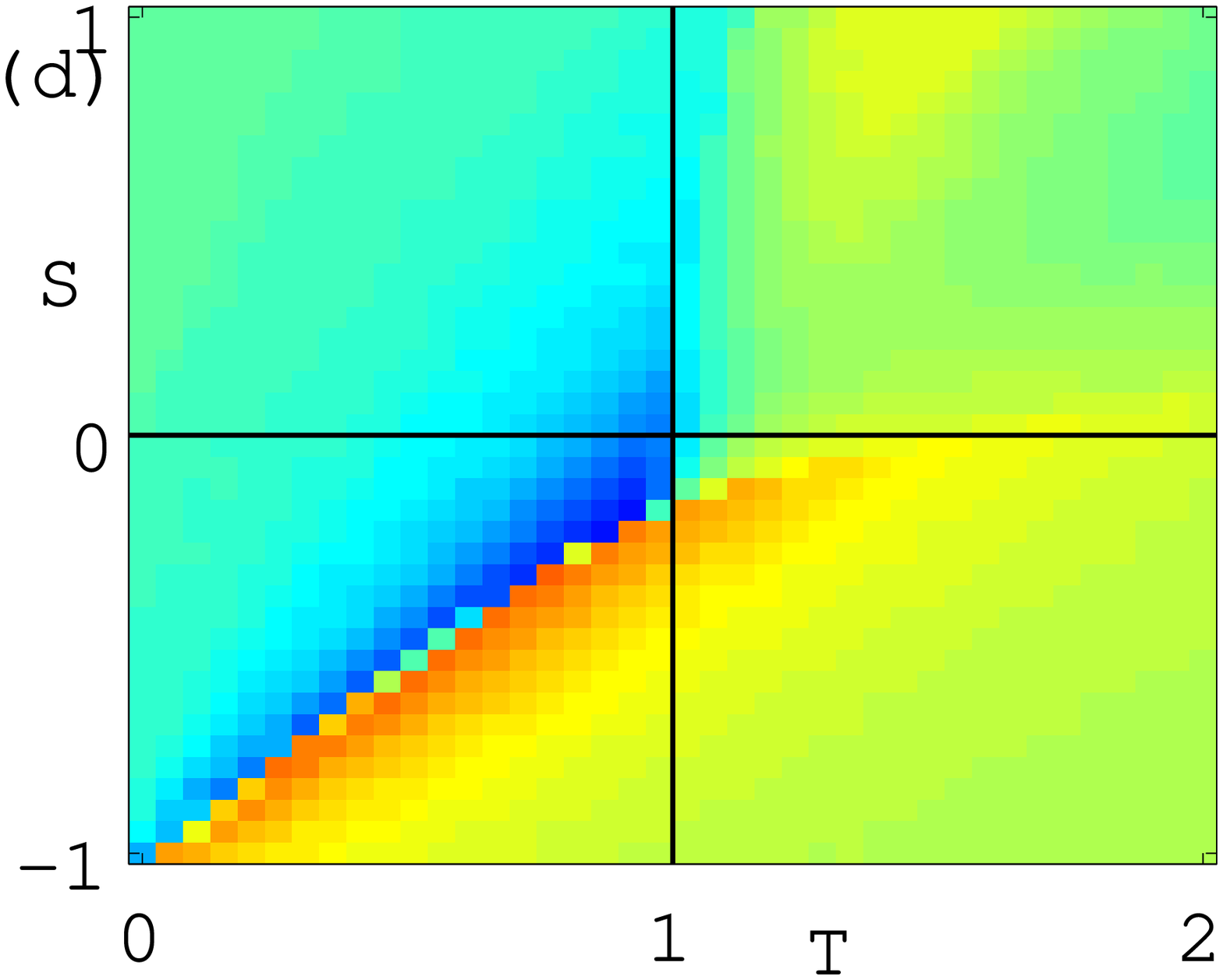}
\caption{}
\label{fig:supp3}
\end{center}
\end{figure}

\end{document}